# Dynamics of neutrino-driven winds: inclusion of accurate weak interaction rates in strong magnetic fields

Men-Quan Liu[1,2]  Zhong-Xiang Wang[1]

[1]  Shanghai Astronomical Observatory, Chinese Academy of Sciences, Shanghai 200030, P.R. China; *liumq@shao.ac.cn*
[2]  Institute of Theoretical Physics, China West Normal University, Nanchong Sichuan, 637002, P.R. China

**Abstract** Solving Newtonian steady-state wind equations with accurate weak interaction rates and magnetic fields (MFs) of young neutron stars considered, we study the dynamics and nucleosynthesis of neutrino-driven winds (NDWs) from proto neutron stars (PNSs). For a typical 1.4 $M_\odot$ PNS model, we find the nucleosynthesis products are closely related to the luminosity of neutrinos and anti-neutrinos. The lower the luminosity is, the larger effect to the NDWs caused by weak interactions and MFs is. At a high anti-neutrino luminosity of typically $8 \times 10^{51}$ erg s$^{-1}$, neutrinos and anti-neutrinos dominate the processes in a NDW and the MFs hardly change the wind's properties. While at a low anti-neutrino luminosity of $10^{51}$ erg s$^{-1}$ at the late stage of a NDW, the mass of product and nucleosynthesis are changed significantly in the strong MFs, they are less important than those in the early stage when the anti-neutrino luminosity is high. Therefore for the most models considered for the NDWs from PNSs, based on our calculations the influences of MFs and the net weak interactions on the nucleosynthesis is not significant.

## 1 INTRODUCTION

A neutrino-driven wind (NDW) from a proto-neutron star (PNS) was firstly proposed by Duncan et al. (1986). It has been regarded as a major candidate site for the $r$-process nucleosynthesis according to observations of metal-poor stars in recent years (Qian 2008). A basic scenario of $r$-process nucleosynsis in a NDW can be simply described as the following (Martinez-Pinedo 2008). Soon after the birth of a PNS, neutrinos and anti-neutrinos are emitted from the surface of the PNS. Because of photodisintegration caused by the shock wave in the supernova explosion, main composition of the circumstellar environment of the PNS is protons, neutrons, electrons and positrons. Main reactions are the absorption and emission of neutrinos and anti-neutrinos by nucleons (so called 'neutrino heat region'). At a certain distance from the PNS, electron fraction $Y_e$ remains constant and $\alpha$ particles are formed. Above this region, other particles, such as $^{12}$C, $^{9}$Be, are successively produced till seed nuclei. Abundant neutrons are captured by the seed nuclei in succession. Neutron-to-seed ratio $\Delta n$ is crucial for building a successful $r$-element pattern, where $\Delta n$ is determined by three essential parameters: electron fraction, entropy, and expansion timescale.

It is difficult to obtain the parameters that fulfill all the nucleosynsis conditions with self-consistency. Detailed analyses for NDWs have been carried out, considering Newtonian and general relativistic hydrodynamics, and physical conditions such as rotating, magnetic fields (MFs) and termination shocks (Kuroda et al. 2008; Metzger et al. 2007; Qian and Woosley 1996; Thompson 2003; Thompson et al. 2001). Among the analyses, the calculations of weak interaction rates are usually treated in an oversimplified way. It is known that young neutron stars (NSs) have a typical MF of $\sim 10^{13}$G, and the surface MF strength of magnetars can be as large as $10^{15}$G (Woods 2006). Strong MFs can potentially change the weak interaction rates significantly. The weak interaction rates decide the electron fraction in the wind, and are very important for the final $r$-process or $\nu p$-process nucleosynthesis. For example, Wanajo et al. (2009) have shown that the puzzle of the excess of $r$-element of A=90 may be solved if $Y_e$ is increased by 1-2%. Recently Arcones et al. (2011) considered the precise weak interaction rates but without MFs. In this paper, we report our work of investigating the NDW in detail considering the accurate weak interaction rates in strong MFs, which is more realistic and practical.

## 2 MODELS

### 2.1 Dynamic equations of NDWs

It has been shown that the steady state is a good approximation for a NDW in the first 20 seconds (Qian and Woosley 1996; Thompson 2003; Thompson et al. 2001). Evolution of the wind is usually obtained by solving the differential equations that include hydrodynamic equations, the equation of state (EoS), and neutrino reaction rates. The dynamical equations of a NDW with the MF pressure are given as follows (Qian and Woosley 1996):

$$4\pi r^2 \rho v = \dot{M}, \tag{1}$$

$$v\frac{dv}{dr} = -\frac{1}{\rho}\frac{dP}{dr} - \frac{GM}{r^2}, \tag{2}$$

$$v\frac{d\varepsilon}{dr} - \frac{v}{\rho^2}P\frac{d\rho}{dr} = \dot{q}, \tag{3}$$

$$v\frac{dY_e}{dr} = \lambda^B_{\nu_e n} + \lambda^B_{e^+ n} - (\lambda^B_{\nu_e n} + \lambda^B_{e^+ n} + \lambda^B_{\bar{\nu}_e p} + \lambda^B_{e^- p})Y_e, \tag{4}$$

$$P = \frac{11\pi^2}{180}\frac{k^4}{(\hbar c)^3}(1 + \frac{30\eta^2}{11\pi^2} + \frac{15\eta^4}{11\pi^4})T^4 + \frac{k}{m_N}\rho T + \frac{B^2}{8\pi}. \tag{5}$$

Here Eq. (1) denotes the mass conservation, where $r$ is the distance from the PNS center to the wind, $v$ is the velocity, and $\dot{M}$ is the outflow rate per second. Eq.(2) denotes the dynamic equilibrium, where $G$ is the gravitational constant and $M$ is the mass of the PNS. Eq.(3) denotes the energy conservation, where $\varepsilon = \frac{11\pi^2}{60\rho}\frac{k^4}{(\hbar c)^3}(1 + \frac{30\eta^2}{11\pi^2} + \frac{15\eta^4}{11\pi^4})T^4 + \frac{3}{2}\frac{k}{m_N}T$ is the internal energy, $\eta \equiv \mu_e/kT$ is the degenerate parameter, and $\mu_e$ is the electron chemical potential. $\dot{q}$ is the total net heating rate for each nucleon (i.e., specific heating rate), and can be divided to two parts: heating and cooling. The Detailed calculations of $\dot{q}$ can be found in Qian and Woosley (1996). Eq.(4) denotes the equilibrium of the composition, that is to say, the change of electrons must be balanced with the variations of the electron fraction, where $\lambda^B_{parameters}$ are a variety of weak interaction rates in MFs. The subscripts of $\lambda^B_{parameters}$ denote the particles participating the reaction. Eq.(5) is the EoS, in which $\hbar$ is the Plank constant, and $c$ is the speed of light. Eq.(5) can be decomposed into three parts: $\frac{\pi^2}{45}\frac{k^4}{(\hbar c)^3}T^4$, $\frac{7\pi^2}{180}\frac{k^4}{(\hbar c)^3}T^4 + \frac{\eta^2}{6}\frac{k^4}{(\hbar c)^3}T^4 + \frac{\eta^4}{12\pi^2}\frac{k^4}{(\hbar c)^3}T^4$, and $B^2/8\pi$. They represent the pressure of photons, electron-positron pairs, and MFs, respectively. In the EoS, we adopted the assumption given by Cooper and Kaplan (2010): the nonradial (e.g., toroidal) component of the MF is large comparing to its radial component. We set the MF as a dipole field: $B(r) = B_i(\frac{R_i}{r})^3$, where $R_i$ and $B_i$ are the radius and the surface MF strength of the PNS. The radius of the PNS was set to be 10 km.

## 2.2 Boundary conditions and numerical techniques

We considered a typical PNS model with a baryonic mass of 1.4 M$_\odot$, obtained in a spherically symmetric simulation of a parameterized 15 M$_\odot$ supernova explosion model. Detailed simulations showed that there are a few $\alpha$ particles appearing at the neutrino sphere, but the number density of $\alpha$ particles is much smaller than those of protons and neutrons (Arcones et al. 2008). As a result, it was reasonable to ignore the influence of $\alpha$ particles on the electron fraction. The initial conditions and neutrino and anti-neutrino luminosities we chose were the same to those in Thopmson et al.(2001): boundary temperature $T_i$=4.3427×10$^{10}$ K, density $\rho_i = 1 \times 10^{13}$ g cm$^{-3}$, and electron fraction $Y_{ei}$=0.03. Two anti-neutrino luminosity models were considered. For anti-neutrino luminosity $L_{\bar{\nu}_e} = 8 \times 10^{51}$ erg s$^{-1}$, we adopted $L_{\nu_e}$=6.15×10$^{51}$ erg s$^{-1}$, $L_{\nu_\mu}$=5.71×10$^{51}$ erg s$^{-1}$, $E_{\nu_e}$=11 MeV, $E_{\bar{\nu}_e}$=14 MeV, $E_{\nu_\mu}$=23 MeV. For anti-neutrino luminosity $L_{\bar{\nu}_e} = 1 \times 10^{51}$ erg s$^{-1}$, we adopted $L_{\nu_e}$=7.69×10$^{50}$ erg s$^{-1}$, $L_{\nu_\mu}$=7.14 ×10$^{50}$ erg s$^{-1}$, $E_{\nu_e}$=6.54 MeV, $E_{\bar{\nu}_e}$=8.32 MeV, $E_{\nu_\mu}$=13.68 MeV.

We solved Eqs.(1)-(5) by using the Runge-Kutta method with the Eqs.(1)-(3) rearranged to be two explicit differential equations of $\rho$ and $T$. It can be noted that if the outflow mass $\dot{M}$ is given, Eqs.(1)-(5) are self-contained, and $\dot{M}$ is kept as a constant during the solving process. Many tries had to be made to find a suitable $\dot{M}$ value for a cross-sonic solution in our solving process. This is because if $\dot{M}$ is too low, the speed of the wind is always lower than the local speed of sound: the wind material will accumulate at a distance not far from the PNS and the wind cannot last. For the opposite case, too large $\dot{M}$ will result in an unphysical infinite acceleration.

Because the parameters, such as $\rho$, vary very quickly near the PNS, the step in solving the equations is required to be sufficiently small in order to keep those parameters change smoothly with a physical meaning. Our initial step was usually less than 10$^{-4}$ cm, while it was adjusted automatically in our code by assuming that the values of all parameters increased in a step were not more than 0.01 of themselves. It is complicated to calculate the electron chemical potentials and weak interaction rates in strong MFs because many integrals and accumulations must be calculated iteratively and the step must be very small as mentioned above. Hence we first obtained the database for the electron chemical potentials and weak interaction rates in the physical environments of the winds, and used their interpolated values during the solving processes of the wind equations.

## 2.3 Influences of MFs

The existence of a MF affects not only the pressure, but also the electron chemical potentials and weak interaction rates. Main influences of the MF on the electron chemical potentials and weak interactions are phase space integrals. In free electron gas, the electron motion perpendicular to a MF is completely determined by the MF and quantized into the well-known Laudau orbitals. The usual sum over electron states (per unit volume) should thus be replaced by the sum of electron states at different energy eigenstates (see e.g., Yuan and Zhang 1998). The details of how to calculate the electron chemical potentials in a strong MF can be found in Zhang et al.(2010). Basically, the number of states for electrons in a MF at momentum interval $[p_z \to p_z + dp_z]$ is proportional to the MF strength, i. e., $\frac{eB}{(2\pi\hbar)^2 c}dp_z$, where $e$ and $p_z$ are the charge and momentum of electrons along the MF respectively. The number of microstates for electrons in the unit volume is equal to the number density of electrons, so for a given number density of electrons, the increase of the MF will reduce the effective integral interval of $dp_z$, and the electron chemical potentials as well. In other words, the MF provides a new state parameter comparing to that without the MF, and increases the state density in the phase space. In the actual numerical calculations, due to the abrupt transitions between the quantized Landau levels at the plane perpendicular to the MF, there may be gurgitation (see Figure 1), but the electron chemical potentials generally decrease under MFs.

Fassio-Canuto first investigated the $\beta$ decay of free neutrons in strong MFs (Fassio-Canuto 1969). Later many authors developed the weak interaction calculations in strong MFs (e.g., Lai and Qian 1991; Luo and Peng1997; Duan et al. 2005). Generally, the effect of a MF on transition matrix element can be ignored comparing to the effect of a MF on phase space. There are mainly four types of weak interactions (including their forward and reverse reactions) in a NDW $\nu_e + n \Leftrightarrow e^- + p$ and $\bar{\nu}_e + p \Leftrightarrow e^+ + n$. For the neutrinos and anti-neutrinos absorption reactions: $\nu_e + n \to e^- + p$ and $\bar{\nu}_e + p \to e^+ + n$, we adopted the method of Lai and Qian (1998), in which both the MF and thermal motion of nucleons were considered. Since the energies of neutrinos and anti-neutrinos are very large (more than 10 MeV for $L_{\bar{\nu}_e} = 8\times 10^{51}$ erg s$^{-1}$), the cross sections change slightly [as shown in Figure 1, 2 and 3 in Lai and Qian (1998)]. As for the electron and positron capture rates $e^- + p \to \nu_e + n$ and $e^+ + n \to \bar{\nu}_e + p$, we used the method of Luo et al. (1997) and Duan et al. (2005). The MF larger than the critical MF ($B_{cr}=4.4\times 10^{13}$ G) can make the rates change dramatically. The lower the positron/electron energy is, the more easily the phase space distribution of positron/electron can be changed.

## 3 RESULTS AND DISCUSSION

The results of the electron chemical potentials as a function of density at different temperatures and MFs are shown in Figure 1. The MF strengths considered were 0, $10^{13}$, $10^{14}$, $10^{15}$G. Comparing to that without a MF, there were obvious changes of the electron chemical potentials for the MF $B=10^{15}$ G at low temperatures (T<$10^{10}$ K) and low densities ($\rho Y_e < 10^9$ g cm$^{-3}$). However when the MF is lower than $10^{15}$ G, such as $10^{13}$, $10^{14}$ G, no significant differences were found. Actually because the curves for such lower MFs nearly coincide with that without a MF, they are not shown in Figure 1. In addition as the temperature near the PNS is very high (more than $10^{10}$ K), the changes of the electron chemical potentials near the surface of the PNS are also not significant.

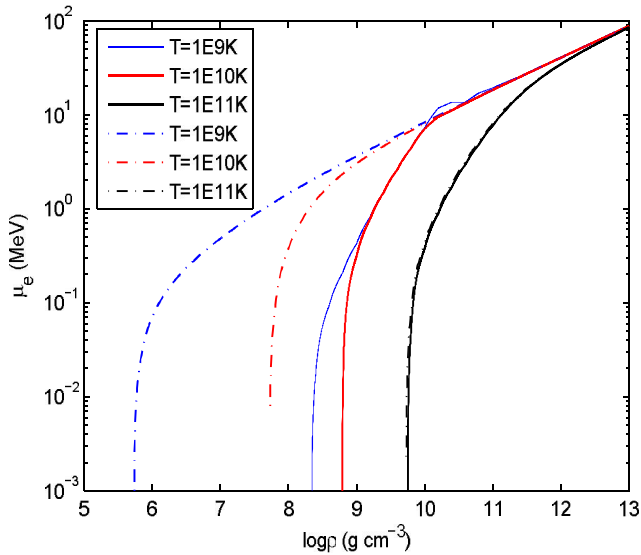

**Fig. 1** The electron chemical potentials (without rest mass) as a function of density at different temperatures. Solid and solid-dotted curves indicate the potentials with $B = 0$ and $B = 10^{15}$ G, respectively.

Referring to Eq.(4) and (5), the electron chemical potentials can change the degenerate parameter and EoS, and the weak interaction rates determine the change of $Y_e$. In Table 1, the results at the critical radius of weak interaction freeze-out are given, at which the final $Y_e$ were also obtained (Martinez-Pinedo 2008). We found that the electron and positron capture rates in the strong MFs can change the $Y_e$ distribution near the surface of the PNS, but hardly change the final $Y_e$ of nucleosynthesis. This is because although the MF larger than the critical MF can make the electron/positron capture rates change largely, the capture rates are generally much less important than neutrino and anti-neutrino absorption rates due to the high energy and luminosity of neutrinos and anti-neutrinos. In addition, the MF strength near the PNS in our dipole model decreases very quickly as the radius increases. For example, for the surface MF strength of the PNS of $10^{15}$ G, it decreases to $3.7\times10^{13}$ G at a radius of 30 km, which is already below the critical MF.

**Table 1** The weak interaction rates at the critical radius of weak interaction freeze-out ($T=10^{10}$ K).

| Model | $\log_{10} B(G)$ | $R_{cr}$(cm) | $\lambda^B_{e^-p}$ (s$^{-1}$) | $\lambda^B_{e^+n}$ (s$^{-1}$) | $\lambda^B_{\bar{\nu}_e n}$(s$^{-1}$) | $\lambda^B_{\nu_e p}$(s$^{-1}$) | $Y_e$ |
|---|---|---|---|---|---|---|---|
| $L_{\bar{\nu}_e,51}=1$ | 13 | 2.16E+06 | 9.15E-02 | 3.59E-01 | 7.21E+00 | 5.81E+00 | 5.62E-01 |
| | 14 | 2.23E+06 | 9.17E-02 | 3.58E-01 | 6.80E+00 | 5.48E+00 | 5.62E-01 |
| | 15 | 3.85E+06 | 9.57E-02 | 3.44E-01 | 2.27E+00 | 1.83E+00 | 5.76E-01 |
| $L_{\bar{\nu}_e,51}=8$ | 13 | 3.85E+06 | 9.39E-02 | 3.50E-01 | 2.67E+01 | 2.85E+01 | 4.86E-01 |
| | 14 | 3.87E+06 | 9.40E-02 | 3.50E-01 | 2.64E+01 | 2.82E+01 | 4.86E-01 |
| | 15 | 4.79E+06 | 9.60E-02 | 3.43E-01 | 1.72E+01 | 1.84E+01 | 4.87E-01 |

*Note 1.* $L_{\bar{\nu}_e,51}$ denotes the luminosity in the units of $10^{51}$ erg s$^{-1}$.

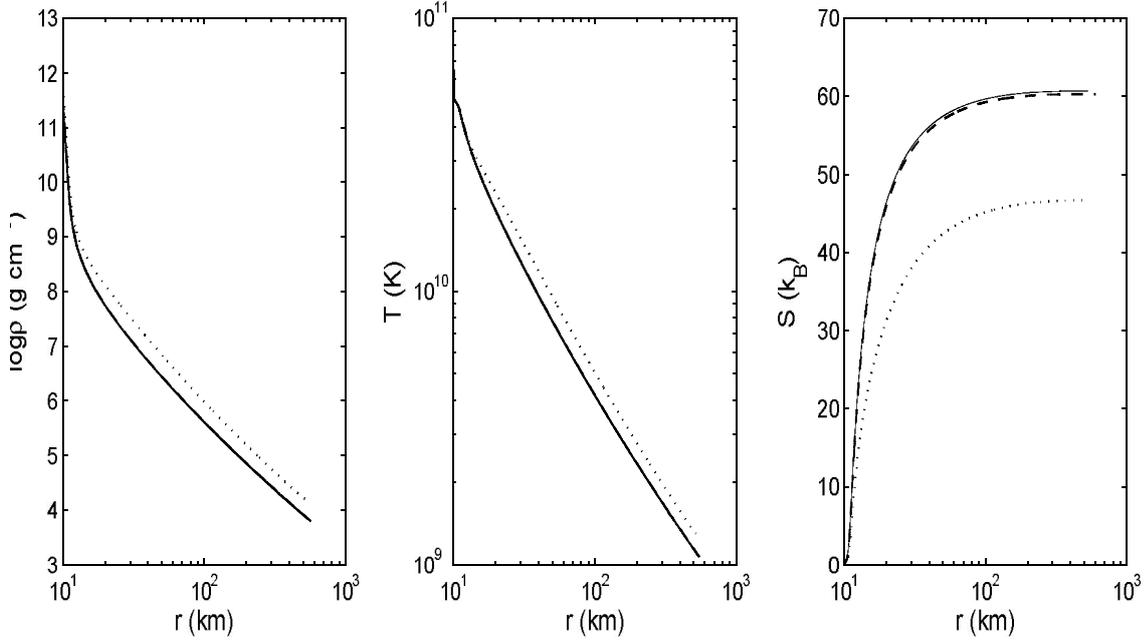

**Fig. 2** The density, temperature and entropy as a function of radius for $L_{\bar{\nu}_e} = 8 \times 10^{51}$erg s$^{-1}$. The solid, dashed and dotted lines indicate the results of $B = 1 \times 10^{13}$, $1\times 10^{14}$ and $1\times 10^{15}$G, respectively.

The obtained properties of the NDW are shown in Figure 2 ($L_{\bar{\nu}_e} = 8 \times 10^{51}$ erg s$^{-1}$) and 3 ($L_{\bar{\nu}_e} = 10^{51}$ erg s$^{-1}$). As can be seen, the larger luminosity of anti-neutrinos is, the smaller the change will be. The reason for this is that the pressure mainly depends on the absorption rates of neutrinos and anti-neutrinos, which are a function of luminosity and average energy of neutrinos and anti-neutrinos. In the high luminosity condition, the MF causes little difference, but for the low luminosity, the case reverses. Although the trends of all parameters, the density, temperature, and entropy, are the same, some differences are clearly visible. The density near the PNS decreases hugely by several orders of magnitude, while drops relatively slowly for the higher anti-neutrino luminosity and stronger MF. Compared to the density, the temperature decreases as a function of the radius much more gently, by a factor $< 5$ at radius $R = 20$ km. The MF and weak interaction rates influence the heating rates, lowering the nucleon entropy especially for the low anti-neutrino luminosity and strong MF case (right panel of Figure 3). Based on the results obtained, we conclude that the density and temperature in our dynamical models have relatively small changes when the MFs are considered, but the variation of nucleon entropy under the influence of a strong MF is significant, which affects the result of nucleosynthesis accordingly.

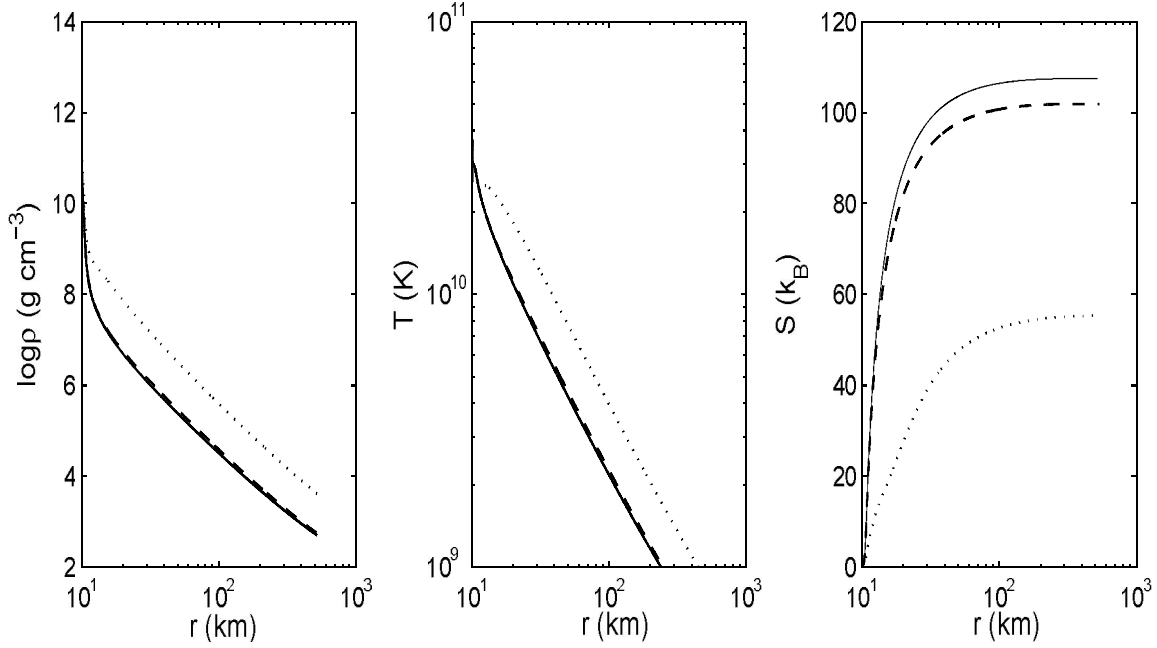

**Fig. 3** The density, temperature and entropy as a function of radius for $L_{\bar{\nu}e}=1\times10^{51}$ erg s$^{-1}$. Notes are the same to Fig.2

In Table 2 a summary of our model calculations is given based on the method described in Roberts et al. (2010). As can be seen, the mass outflow rate increases with the anti-neutrino luminosity, and the MF enhances the total mass of the NDW. $r$-element distribution patterns depend on the entropy, expansion timescale and electron fraction. For the nucleon entropy, we found that it changes to be smaller under the strong MFs, which is caused by the difference of heating rate and heating timescale. For the expansion timescale $\tau_d$, we used the definition of Metzger et al. (2007), and our results show that in the MF $B=10^{15}$ G, the expansion timescale is much larger than the others. As for the electron fraction, they remain nearly the same even under the ultra-strong MF for high anti-neutrino luminosity.

**Table 2** The nucleosynthesis parameters resulting from our models.

| Model | $\log_{10}B(G)$ | $\dot{M}(M_\odot\,s^{-1})$ | $S(k)$ | $\tau_d$(ms) | $Y_e$ | $Y_s$ | $\Delta_n$ |
|---|---|---|---|---|---|---|---|
| $L_{\bar{\nu}_e,51}=8$ | 13 | 2.40E-4 | 60.66 | 10.55 | 0.486 | 2.76E-3 | 10.13 |
|  | 14 | 2.41E-4 | 60.28 | 10.73 | 0.486 | 2.77E-3 | 10.11 |
|  | 15 | 3.53E-4 | 46.70 | 18.22 | 0.487 | 2.60E-3 | 10.00 |
| $L_{\bar{\nu}_e,51}=1$ | 13 | 1.58E-6 | 107.50 | 123.62 | 0.562 | 1.86E-3 | 1.815 |
|  | 14 | 1.65E-6 | 101.90 | 143.76 | 0.562 | 2.29E-3 | 1.569 |
|  | 15 | 5.47E-6 | 55.3 | 469.86 | 0.576 | 8.30E-3 | 0.440 |

For model $L_{\bar{\nu}_e,51}=8$, $Y_e=0.486$, which implies neutrino-rich in the wind. $r$-process nucleosynthesis begins after the charged particle reactions, and the total quantity of nucleosynthesis is determined by the outflow mass. It can be seen from Table 2 that the mass outflow rate with the MF of B=$10^{15}$ G is increased by about 1/3, and the changes of $S$ and $\tau_d$ are also significant for the case B=$10^{15}$ G. However the differences in seed nuclei abundance $Y_s$ and neutron-to-seed ratio $\Delta_n$ at different MFs are very small.

The reasons for this are the following. In the stage of charged particle reactions, high entropy will effectively reduce the production rate of $^{12}$C, because high entropy values means more photons and thus photodisintegration becomes faster. We note that since the nuclear reaction that generates $^{12}$C is the slowest in the process of generating seed nuclei, the number of $^{12}$C is approximately equal to the number of seed nuclei. On the other hand the slower the expansion is, the more slowly the density reduces. Since $^{12}$C is produced by a three-body reaction, whose rate is proportional to the square of density, the increased expansion timescale makes the products of $^{12}$C increased (Martinez-Pinedo 2008). For our results, we found that the entropy decreases significantly, and the expansion timescale also increases for the strongest MF case. Combing these two factors, for the same $Y_e$, the seed nuclei abundance $Y_s$ should be increased. However the differences of the obtained $Y_s$ are very small (Table 2). This is because the entropy is not large enough to affect the $Y_s$, as $Y_s \approx (0.1 - 0.2Y_e) \times [1 - \exp(-8 \times 10^8 \tau_d S_f^{-3} Y_e^3)]$ (Roberts et al. 2010). Only when the entropy is greater than 70, the change of $Y_s$ is sensitive to the parameters. The abundance of free neutrons to seed nuclei is closely dependant on $Y_s$, and thus $\Delta_n$ also had small changes.

For model $L_{\bar{\nu}_e,51}$=1, $Y_e$>0.5, which means proton-rich in the wind. For convenience of discussion, the nucleosyhthesis can be separated into two phases. During the first phase, the rapid proton capture process similar to that in Type I X-ray bursts will take place and end at the heavy nuclei with a long half-life such as $^{64}$Ge. These nuclei are prohibited to capture protons and are the seed nuclei in the second phase. During the second phase, a large number of the anti-neutrinos are absorbed continuously by protons and produce neutrons (the so called $\nu$p process). The number density of neutrons can reach $10^{14-15}$ cm$^{-3}$. These neutrons, not affected by the Coulomb repulsion energy, are easily captured by the seed nuclei through a series of $(n,p)(p,\gamma)$ reactions to produce heavier nuclei, effectively through nuclei with a long half-life (Frohlich et al. 2006). Therefore in a NDW, the abundance of free neutrons is essential to the final products. Based on the rough considerations (Pruet 2006), the relative abundance of nucleosynthesis is proportional to $e^{-\Delta_n}$, and the relative abundances were estimated to be 0.1665, 0.2122 and 0.6608 for $B = 10^{13}, 10^{14}$, and $10^{15}$G in our model, respectively. This implies that the yield of $\nu$p-process is significantly increased in the strong MFs.

In this paper, using the classic calculation methods of weak interactions and electron chemical potentials, both the accurate weak interaction rates and influences of strong MFs were considered for the NDW from a PNS. We also estimated the nucleosynthesis for the two cases: neutron-rich and proton-rich, by employing analytic methods. We found that neutrino and anti-neutrino luminosities are important factors to the results. For a high luminosity case the influence of a strong MF is small because the effect of neutrinos dominates. We note that at the late stage of a NDW when the neutrino luminosity will be relatively low, the nucleosynthesis of $\nu$p-process in strong MFs can change significantly. For example, if the model luminosity of anti-neutrino is $10^{50}$ erg s$^{-1}$, the entropies will be 199 $k$ and 77 $k$ for B=$10^{13}$ G and $10^{15}$ G respectively. The difference is 2.6 times. The outflow rates will be $7.0 \times 10^{-9}$ M$_\odot$ s$^{-1}$ for B=$10^{13}$ G and $3.55 \times 10^{-8}$ M$_\odot$ s$^{-1}$ for B=$10^{15}$ G, which means the yield in the stronger MF is about 5 times larger than that in the lower one. However because the yield at this late stage is much less than that in the early stage when the neutrino luminosity is high, the influence of MFs to the total yield is not significant.

**Acknowledgements** This work is supported by the National Natural Science Foundation of China (Grant Nos.11073042,11273020), National Basic Research Program of China (973 Project 2009CB824800), China Postdoctoral Science Foundation funded project(2012T50446), Youth Fund of Sichuan Provincial Education Department (10ZC014,2009ZB087), and Science and Technological Foundation of CWNU. ZW is a Research Fellow of the One-Hundred-Talents project of Chinese Academy of Sciences.

**References**

Arcones, A., & Martinez-Pinedo, G. 2011, Phys Rev C, 83, 045809
Arcones, A., Martinez-Pinedo, G., O'Connor, E., Schwenk, A., Janka, H. T., Horowitz, C. J., & Langanke, K. 2008, Phys Rev C, 78, 5806
Cooper, R. L., & Kaplan, D. L. 2010, Astrophys J Lett, 708, L80
Duan, H., & Qian, Y.-Z. 2005, Phys Rev D, 72, 23005
Duncan, R. C., Shapiro, S. L., & Wasserman, I. 1986, Astrophys J, 309, 141
Fassio-Canuto. 1969, Physics Review, 187, 2138
Frohlich, C., et al. 2006, Phys Rev Lett, 96, 142502
Kuroda, T., Wanajo, S., & Nomoto, K. 2008, Astrophys J, 672, 1068
Lai, D., & Qian, Y.-Z. 1998, The Astrophysical Journal, 505, 844
Lai, D., & Shapiro, S. L. 1991, The Astrophysical Journal, 383, 745
Luo, Z.-Q., & Peng, Q.-H. 1997, Chinese Astronomy and Astrophysics, 21, 254
Martinez-Pinedo, G. 2008, Eur Phys J-Spec Top, 156, 123
Metzger, B. D., Thompson, T. A., & Quataert, E. 2007, Astrophys J, 659, 561
Qian, Y. Z. 2008, in Proceedings of Science the 10th Symposium on Nuclei in the Cosmos (Michigan,USA), 1
Qian, Y. Z., & Woosley, S. E. 1996, Astrophys J, 471, 331
Roberts, L. F., Woosley, S. E., & Hoffman, R. D. 2010, Astrophys J, 722, 954
Thompson, T. A. 2003, Astrophys J, 585, L33
Thompson, T. A., Burrows, A., & Meyer, B. S. 2001, Astrophys J, 562, 887
Woods, P. M., & Thompson, C. 2006, in Compact Stellar X-ray Sources, ed. L. Walter (Cambridge University Press)
Wanajo, S., Nomoto, K., Janka, H. T., Kitaura, F. S., & Muller, B. 2009, Astrophys J, 695, 208
Yuan, Y. F., & Zhang, J. L. 1998, Astronomy and Astrophysics, 335, 969
Zhang, J., Wang, S. F., & Liu, M. Q. 2010, Int J Mod Phys E, 19, 437